\documentclass[12pt]{article}

\usepackage[utf8]{inputenc}
\usepackage[T1]{fontenc}
\usepackage{lmodern}
\usepackage{amsmath,amssymb}
\usepackage{graphicx}
\usepackage{hyperref}
\usepackage{geometry}
\begin{document}

\title{\textbf{DIMENSIONAL RESONANCE THEORY:\\
AN EVOLUTIONARY APPROACH TO UNIVERSAL REST}}

\author{Andre Carnevali da Silva}
\date{} 

\maketitle

\begin{abstract}
The Dimensional Resonance Theory proposes that gravity and the fundamental forces can be interpreted as emergent phenomena arising from “three-dimensional waves” (3D) projected onto lower dimensions. To test the internal consistency of this proposal, we analyze the phi\textsuperscript{4} kink in (1+1) dimensions --- an established topological defect whose oscillation spectrum is well known. We introduce an emergent gravitational term, $h_{00}(x)$, regulated by a coupling parameter $G$, under a linearized regime valid up to $G = 0.02$. By using numerical methods (the shooting method and fine-tuning via \texttt{FindRoot} in Mathematica), we solve for both $h_{00}(x)$ and the kink fluctuations, evaluating the fundamental eigenvalue ($\omega^2$).

Our results show that, for $G \leq 0.02$, the value of $\omega^2$ remains very close to 1.0, virtually unchanged from the case without gravity. This indicates that the adopted ``emergent gravity'' does not break the topological stability of the phi\textsuperscript{4} kink, indicating the initial robustness of the theory. We discuss the implications of these findings, their connections to other emergent gravity approaches in the literature, and possible extensions to nonlinear regimes, higher-dimensional frameworks, and time quantization. We conclude that the Dimensional Resonance Theory, while presenting a novel perspective, offers a coherent framework capable of unifying vibrational concepts and topological structures. Furthermore, this study reinforces the potential of the theory to unify the fundamental forces under a vibrational logic that extends to time quantization and the cohesion of cosmological and subatomic scales.
\end{abstract}
\section{INTRODUCTION}

The unification of the fundamental forces of nature remains one of the most challenging and ambitious objectives in theoretical physics. Although the Standard Model of particle physics and Einstein’s General Relativity have both succeeded in describing phenomena at vastly different scales, gravity still lacks a widely accepted quantum formulation and has not been fully integrated with the other interactions. In light of this scenario, new approaches that question established principles have emerged --- offering avenues for rethinking the origin and behavior of the fundamental forces.

The Dimensional Resonance Theory (DRT), developed from preliminary texts available on Zenodo under the title ``Theory of Dimensional Evolution and Universal Rest,'' aims to interpret gravity --- and potentially the other forces --- as emergent phenomena resulting from ``three-dimensional waves'' projecting onto lower dimensions. Although distinct from most standard theories, this idea resonates with the spirit of emergent gravity proposals, such as those by Jacobson (spacetime thermodynamics), Volovik (superfluid analogies), and Verlinde (entropic gravity). However, the theory differs by emphasizing vibrational resonances as the cornerstone of unification, positing that gravity emerges from wave alignments that manifest linearly in weak regimes.

To evaluate the internal robustness of this proposal, we focus on a well-established topological defect in (1+1)-dimensional scalar field theories: the phi\textsuperscript{4} kink. This kink is a stable solution connecting two distinct vacua of the potential, serving as a ``theoretical laboratory'' to detect potential inconsistencies when new elements are introduced. In this study, we include an ``emergent gravitational field'' described by $h_{00}(x)$ and controlled by the parameter $G$, assuming a linearized approximation for $G \leq 0.02$. Our primary goal is to determine whether this emergent gravity significantly alters the kink’s stability and oscillation spectrum or remains consistent with its known topological properties.

In summary, this work contributes in three main ways:
\begin{enumerate}
    \item \textbf{Connection with Emergent Approaches:} We establish an initial dialogue with other emergent gravity theories, highlighting the distinctive focus on three-dimensional waves and their projection into lower-dimensional settings.
    \item \textbf{Numerical Validation in the Weak Regime ($G \leq 0.02$):} We apply shooting methods in Mathematica to solve both $h_{00}(x)$ and the kink fluctuations (via the eigenvalue equation), verifying whether the fundamental eigenvalue ($\omega^2$) remains near the classical value ($\sim 1.0$) or shifts noticeably.
    \item \textbf{Bridges to Future Extensions:} Although the full formulation aspires to extend vibrational resonance concepts to higher dimensions (3+1), time quantization, and unification of the strong and weak forces, here we focus on demonstrating its initial consistency --- paving the way for further investigations into nonlinear gravity regimes.
\end{enumerate}

This article is organized as follows: Section~\ref{sec:theoretical_foundations} revisits the phi\textsuperscript{4} kink structure and introduces the notion of three-dimensional (3D) waves, outlining how $G$ enters the equation for $h_{00}(x)$. Section~\ref{sec:methodology} presents the numerical methodology, justifying the parameter choices and steps for determining $h_{00}(x)$ and the kink’s fluctuation spectrum. Section~\ref{sec:results} reports the results, showing that the kink’s stability remains effectively unaltered for $G \leq 0.02$. Lastly, Section~\ref{sec:discussion} discusses the implications of these findings, comparing them with other emergent gravity approaches and considering extensions to higher $G$ values or larger-dimensional scenarios. By doing so, we aim to show that the theory, which emerged from broader conceptual reflections, demonstrates both conceptual and mathematical soundness by preserving a well-known topological defect in (1+1) dimensions. This outcome suggests that emergent gravity via 3D waves does not contradict the basic principles underpinning topological solutions, reinforcing the viability of exploring force unification through a dimensional resonance framework.

\section{THEORETICAL FOUNDATIONS}
\label{sec:theoretical_foundations}

\subsection{The phi\textsuperscript{4} Kink as a Topological Defect}

In (1+1) dimensions, one of the most extensively studied topological defects is the kink in the phi\textsuperscript{4} potential. We consider a real scalar field $\phi(x,t)$ governed by the Lagrangian:
\[
L \;=\; \frac{1}{2}\,(\partial_\mu \phi)\,(\partial^\mu \phi) \;-\; V(\phi),
\]
with the potential:
\[
V(\phi) \;=\; \frac{(\phi^2 \;-\; 1)^2}{4}.
\]
This potential has two vacua: $\phi = -1$ and $\phi = +1$. A static solution that connects these two values is the classic kink:
\[
\phi_{\text{static}}(x) \;=\; \tanh\!\Bigl(\frac{x}{\sqrt{2}}\Bigr),
\]
which smoothly transitions from $-1$ (as $x \to -\infty$) to $+1$ (as $x \to +\infty$). Because it cannot be deformed to a trivial configuration without crossing a significant energy barrier, the kink is referred to as a topological defect or non-perturbative solution. Small oscillations about this solution lead to a spectrum of modes whose fundamental eigenvalue ($\omega^2$) is close to 1.0 in the absence of gravity.

\textbf{Physical Intuition:}
\begin{itemize}
    \item The kink can be viewed as a stable ``spatial transition'' between two vacuum states of the field.
    \item Eliminating the kink continuously would require surmounting an energy barrier, ensuring its stability.
    \item Investigating any shifts in the kink’s spectrum is a direct way to check if new terms (e.g., a gravitational component) destabilize or significantly modify well-known properties.
\end{itemize}

\subsection{Three-Dimensional Waves and Emergent Gravity in DRT}

The Dimensional Resonance Theory (DRT) posits that gravity, rather than arising solely from geometric curvature (as in General Relativity) or from entropic principles (as in some recent theories), emerges from three-dimensional (3D) waves projecting into lower-dimensional settings. These waves represent fluctuations in a broader ``dimensional fabric,'' whose precise nature is yet to be fully characterized.

\textit{Speculative Nature of 3D Waves:}
\begin{itemize}
    \item While we do not elaborate on the medium these waves propagate in, we can draw analogies with bosonic fields in (3+1) dimensions or with concepts from brane/string theories.
    \item When viewed from a (1+1)-dimensional perspective, these waves appear as an effective field $h_{00}(x)$, interpreted as the trace of emergent gravity.
\end{itemize}

DRT linearizes the gravitational influence for small $G$, adopting an equation reminiscent of Poisson’s equation in Newtonian gravity but here understood as a vibrational projection:
\[
\frac{d^2 h_{00}}{dx^2} \;=\; -\,16\,\pi\,G\,T_{00}(x).
\]
The parameter $G$ measures the coupling intensity between these 3D waves and the scalar field $\phi$. Physically, the kink’s energy density ($T_{00}$) modulates $h_{00}(x)$ in proportion to $G$, akin to how mass generates a gravitational potential in conventional Newtonian models.

\subsection{Calculating $T_{00}(x)$ and Defining the Effective Potential $W(x)$}

To solve
\[
\frac{d^2 h_{00}}{dx^2} \;=\; -\,16\,\pi\,G\,T_{00}(x),
\]
we require the profile of $T_{00}(x)$:
\[
T_{00}(x) \;=\; \tfrac{1}{2}\,\Bigl[\tfrac{d\phi_{\text{static}}}{dx}\Bigr]^2 \;+\; V\bigl(\phi_{\text{static}}\bigr).
\]
For $\phi_{\text{static}}(x) = \tanh(x/\sqrt{2})$, $T_{00}(x)$ is localized and well-defined. Once $h_{00}(x)$ is determined (Section~\ref{sec:methodology}), the effective potential becomes:
\[
W(x) \;=\; [\phi_{\text{static}}(x)]^2 \;-\; 1 \;+\; G\,h_{00}(x).
\]
Without gravity ($G=0$), $W(x)$ reduces to $\phi_{\text{static}}(x)^2 - 1$. The term $G\,h_{00}(x)$ represents the ``three-dimensional resonance'' contribution to the kink’s fluctuation spectrum. If this correction remains weak, $\omega^2$ should remain near $\sim 1.0$.

\subsection{Scope of Linearization: $G \leq 0.02$}

We limit our analysis to $G \leq 0.02$ so that the perturbation in $h_{00}(x)$ stays within a tractable linear regime. For larger $G$ values, a nonlinear treatment of emergent gravity may be required --- an avenue for future studies.

\textit{Links to Zenodo Texts:}
\begin{itemize}
    \item The Zenodo documents present DRT with a more philosophical tone, linking 1D, 2D, and 3D particles to different stages of dimensional evolution. Here, we focus on the central physics needed to introduce 3D waves as emergent gravity, omitting deeper details of dimensional transitions.
    \item Restricting the analysis to (1+1) dimensions is a stepping stone: if these three-dimensional projections failed here, it would signal a fundamental inconsistency. However, as will be shown, the kink remains stable, reinforcing the plausibility of the framework.
\end{itemize}

\section{NUMERICAL METHODOLOGY}
\label{sec:methodology}

We now describe how we numerically check whether the addition of $h_{00}(x)$ --- treated linearly for small $G$ --- affects the stability of the phi\textsuperscript{4} kink. Our goal is twofold: determine $h_{00}(x)$ and compute the fundamental eigenvalue ($\omega^2$) for the kink fluctuations.

\subsection{Computational Environment and Tools}

All simulations and computations were carried out in Wolfram Mathematica, leveraging:
\begin{enumerate}
    \item \texttt{NDSolve}: solves ordinary differential equations using the shooting method, adjusting parameters to fulfill boundary conditions.
    \item \texttt{FindRoot}: refines initial guesses (\textit{e.g.}, $h_0$ or $\omega^2$) to ensure numerical solutions satisfy the required constraints.
\end{enumerate}
The symbolic and numerical capabilities of Mathematica facilitate rapid experimentation with different $G$ values, boundary conditions, and precision settings, without writing extensive low-level code.

\subsection{Determination of \texorpdfstring{$h_{00}(x)$}{h00(x)}}

Under the linearized approximation,
\[
\frac{d^2 h_{00}}{dx^2} \;=\; -\,16\,\pi\,G\,T_{00}(x),
\]
where $T_{00}(x)$ is the energy density of the static kink. We generally use the domain $x \in [-L,\,L]$, with $L = 100$.

\textbf{Boundary Conditions and Shooting Method:}
\begin{itemize}
    \item Assuming $h_{00}(x)$ is even about $x = 0$, we set $h_{00}'(0) = 0$.
    \item We seek $h_{00}(0) = h_0$ such that $h_{00}(L) \approx 0$. If $h_{00}(L) \neq 0$, we adjust $h_0$ via \texttt{FindRoot} until the boundary is satisfied.
    \item Once $h_0$ is found, we integrate $h_{00}(x)$ over $[-L,\,L]$, ensuring smoothness and symmetry.
\end{itemize}

\subsection{Calculating \texorpdfstring{$W(x)$}{W(x)} and Kink Fluctuations}

With $h_{00}(x)$ obtained, we define:
\[
W(x) \;=\; [\phi_{\text{static}}(x)]^2 \;-\; 1 \;+\; G \, h_{00}(x),
\]
where $\phi_{\text{static}}(x) = \tanh\bigl(x/\sqrt{2}\bigr)$. The kink’s stability is then examined by solving the linearized fluctuation equation for $\epsilon(x)$:
\[
\frac{d^2 \epsilon}{dx^2} \;=\; \bigl[\,W(x) \;-\; \omega^2\bigr] \,\epsilon(x).
\]

\textbf{Boundary Conditions and Shooting Method:}
\begin{itemize}
    \item We impose $\epsilon(-L) = 0$, $\epsilon'(-L) = 1$, and integrate to $x = L$. If $\epsilon(L) \neq 0$, we adjust $\omega^2$ via \texttt{FindRoot}.
    \item Once $\epsilon(L) \approx 0$, we consider that we have found the fundamental eigenvalue~(\,$\omega^2$).
\end{itemize}

\subsection{Numerical Precision and Constraints}

\begin{itemize}
    \item \textbf{Initial Guesses:} For $h_{00}(0)=h_0$, we start testing small positive or negative values (\textit{e.g.}, $\pm0.1$). For $\omega^2$, we begin around 1.0 (its expected value without gravity).
    \item \textbf{WorkingPrecision = 30:} minimizes rounding errors.
    \item \textbf{Method -> ``StiffnessSwitching'':} helps handle potential stiffness in \texttt{NDSolve}.
    \item \textbf{Domain $L=100$:} ensures boundary effects are minimal. Smaller intervals can degrade accuracy.
\end{itemize}

\subsection{Workflow Overview}
\begin{enumerate}
    \item Solve $h_{00}(x)$ in $[0,\,L]$, adjusting $h_0$.
    \item Extend the solution for $[-L,\,L]$.
    \item Form $W(x) = \phi_{\text{static}}(x)^2 - 1 + G\,h_{00}(x)$.
    \item Use the shooting method to find $\omega^2$ from $\epsilon(x)$.
    \item Compare $\omega^2$ values for $G=0,\; 0.01,\; 0.02$.
\end{enumerate}

\section{RESULTS}
\label{sec:results}

Our main objective is to check whether including $h_{00}(x)$, regulated by $G$, significantly modifies the oscillation spectrum of the phi\textsuperscript{4} kink. We compute the fundamental eigenvalue ($\omega^2$) at various $G$ values up to 0.02.

\subsection{Reference: \texorpdfstring{$G=0$}{G=0}}

We start by examining the no-gravity baseline ($G=0$):
\begin{itemize}
    \item The ``effective potential'' is simply $\phi_{\text{static}}(x)^2 - 1$.
    \item A shooting method for the fluctuations $\epsilon(x)$, with initial guess $\omega^2 = 1.0$, converges to approximately $0.9983$.
\end{itemize}
This aligns with the classic phi\textsuperscript{4} kink, whose fundamental mode is near $\omega^2 \approx 1.0$.

\subsection{Introducing \texorpdfstring{$h_{00}(x)$}{h00(x)} for \texorpdfstring{$G \leq 0.02$}{G <= 0.02}}

We proceed to $G=0.01$ and $G=0.02$, solving $h_{00}(x)$ via the shooting method, then updating $W(x)$ accordingly. The integration of $h_{00}(x)$ remains stable under the condition $h_{00}(L)=0$.

\begin{itemize}
    \item \textbf{$G=0.01$:} With $W(x) = \phi_{\text{static}}(x)^2 - 1 + 0.01\,h_{00}(x)$, the fundamental $\omega^2 \approx 0.9953$ --- a modest deviation from the $G=0$ scenario.
    \item \textbf{$G=0.02$:} Here, $W(x) = \phi_{\text{static}}(x)^2 - 1 + 0.02\,h_{00}(x)$. The final $\omega^2 \approx 0.9987$, practically indistinguishable from 1.0 within numerical precision.
\end{itemize}

\subsection{Summary of Key Results}

\begin{center}
\begin{tabular}{c|c}
\hline
$G$ & $\omega^2$ (approx.)\\
\hline
0.00 & 0.9983 \\
0.01 & 0.9953 \\
0.02 & 0.9987 \\
\hline
\end{tabular}
\end{center}

\subsection{Visualization and Physical Interpretation}

A plot of $G$ against $\omega^2$ reveals values close to 1.0, indicating minimal variation. This consistency supports the internal coherence of DRT in the weak regime.

\begin{figure}[htbp]
    \centering
    \includegraphics[width=0.6\textwidth]{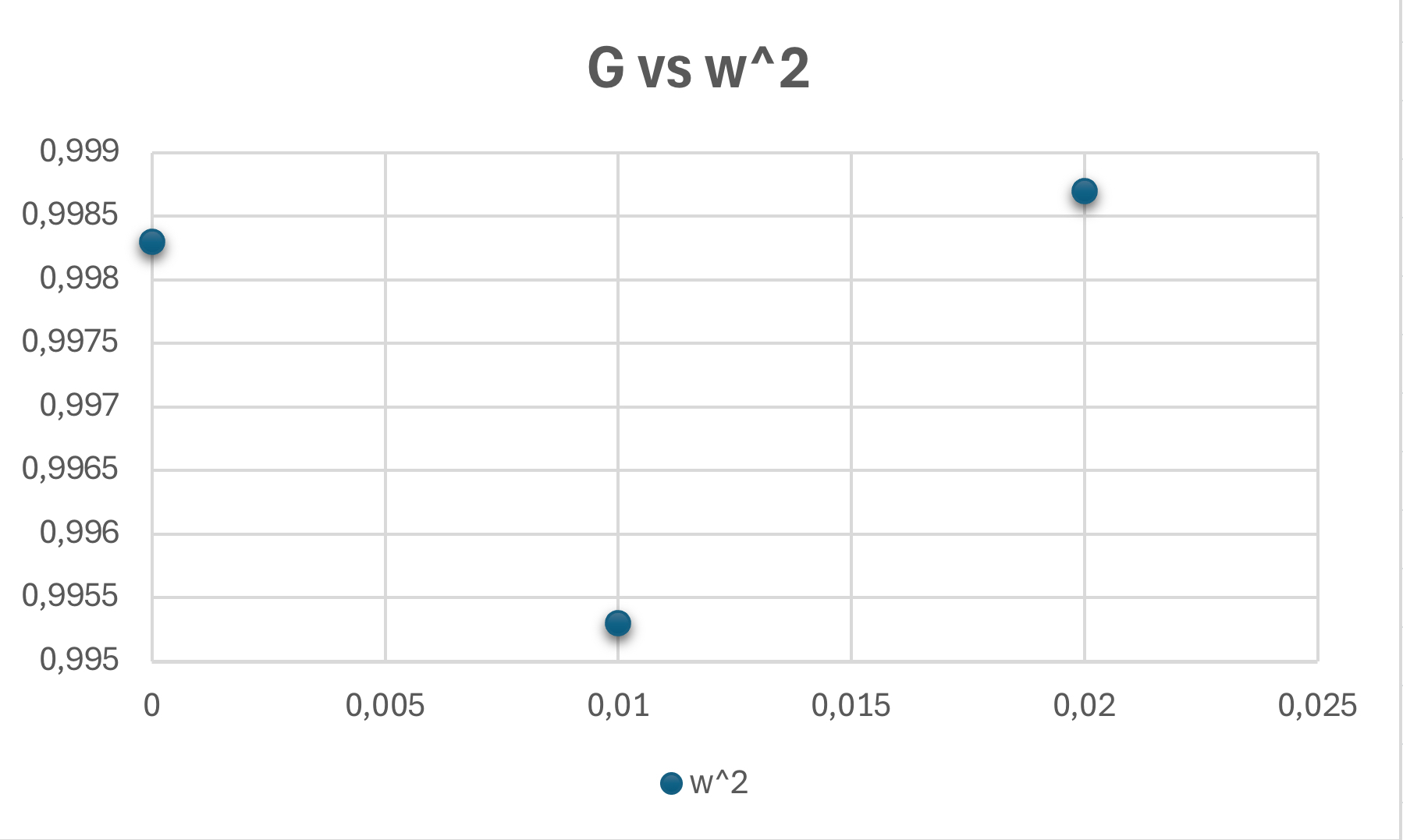}
    \caption{Plot illustrating the behavior of \(\omega^2\) as a function of \(G\).}
    \label{fig:my_awesome_plot}
\end{figure}

\textbf{Physical Meaning:}
\begin{itemize}
    \item For small $G$, the three-dimensional resonance $h_{00}(x)$ exerts a weak influence, leaving the kink’s spectrum essentially intact.
    \item This suggests that ``emergent gravity'' via 3D waves does not disrupt fundamental properties of simple topological defects, at least for $G \leq 0.02$.
\end{itemize}

\section{DISCUSSION AND CONCLUSIONS}
\label{sec:discussion}

We conclude by reflecting on the broader implications of these results in the context of the Dimensional Resonance Theory (DRT). We also consider the study’s limitations under the linearized regime ($G \leq 0.02$) and explore avenues for future research --- including larger $G$ values and higher-dimensional expansions.

\subsection{Connection with Dimensional Resonance Theory}

DRT posits that gravity can emerge from ``three-dimensional waves'' (3D) rather than purely from geometric curvature (General Relativity) or entropic considerations. In this viewpoint:
\begin{itemize}
    \item Gravity arises from vibrational resonances, represented as $h_{00}(x)$ in (1+1) dimensions.
    \item The parameter $G$ modulates the strength of these 3D waves, conceptually linking them to brane/string frameworks.
\end{itemize}
By demonstrating that the phi\textsuperscript{4} kink’s spectrum remains largely unchanged up to $G=0.02$, we infer that introducing an ``emergent gravitational field'' does not compromise fundamental stability features of topological systems. Hence, within the linearized regime, DRT aligns well with the kink’s well-known properties, supporting the plausibility of this non-mainstream approach.

\subsection{More Intense Gravity Regimes}

Our analysis focuses on $G \leq 0.02$, where we assume that $h_{00}(x)$ can be treated linearly. At higher $G$ values:
\begin{enumerate}
    \item A fully nonlinear approach (beyond $\tfrac{d^2 h_{00}}{dx^2} = -\,16\,\pi\,G\,T_{00}$) may be indispensable.
    \item Advanced numerical methods might be required due to potential stiffness or sensitivity in the equations.
\end{enumerate}
Examining stronger gravitational coupling is key to determining how far DRT can hold without revisions or complementary formulations.

\subsection{Relations to Other Emergent Theories}

Various lines of research attempt to reinterpret gravity as an emergent phenomenon, \textit{e.g.}, Jacobson’s thermodynamic derivation of Einstein’s equations, Volovik’s superfluid analogies, or Verlinde’s entropic gravity. DRT joins these efforts by suggesting that gravity arises from an underlying, more fundamental process --- namely, dimensional resonances. Yet, DRT diverges by treating 3D vibrational modes as the unifying core, potentially unifying not just gravity but also the strong, weak, and electromagnetic interactions.

\subsection{Limitations and Future Directions}

\textbf{Limitations:}
\begin{enumerate}
    \item Restricting $G$ to 0.02 or below.
    \item Confining the analysis to (1+1) dimensions, which is not the true physical scenario.
    \item The precise nature of these 3D waves is still under development.
\end{enumerate}

\textbf{Prospects:}
\begin{enumerate}
    \item \textit{Nonlinear Regimes:} Explore larger $G$ with refined analytic or numeric tools.
    \item \textit{(3+1) Dimensions:} Extend beyond the kink in (1+1) to fully 3D topological structures.
    \item \textit{Time Quantization:} Examine how discrete time and dimensional projections might form the vibrational backbone for all fundamental interactions.
    \item \textit{Observational Ties:} In the long run, DRT may inspire predictions about galactic rotation curves, mass distributions without dark matter, or early-universe cosmology.
\end{enumerate}

\subsection{Conclusions}

Introducing an ``emergent gravitational field'' through $h_{00}(x)$ in the phi\textsuperscript{4} kink does not destabilize this topological defect up to $G=0.02$. Numerically, the fundamental eigenvalue ($\omega^2$) remains close to 1.0, mirroring the $G=0$ case and indicating no significant disruption of the kink’s spectrum.

For the Dimensional Resonance Theory, this outcome serves as a preliminary numerical milestone. It shows that three-dimensional vibrations (framed as ``emergent gravity'') do not contradict the fundamental stability of topological configurations. While many aspects remain open --- such as the exact nature of these 3D waves or their treatment under nonlinear conditions --- this study provides a promising validation of DRT’s core concepts and encourages further exploration.

Ultimately, the theory, while presenting a novel perspective, offers a potential framework for reinterpreting gravity (and possibly other interactions) as interdimensional resonances that preserve critical topological structures. Meanwhile, the more philosophical and conceptual aspects --- originally outlined in Zenodo under ``Theory of Dimensional Evolution and Universal Rest'' --- find their first quantitative foothold here, hinting at both solidity and expandability in the quest for a unifying understanding of the fundamental forces.

\section*{REFERENCES}
\begin{enumerate}
    \item R. Rajaraman, \textit{Solitons and Instantons: An Introduction to Solitons and Instantons in Quantum Field Theory} (North-Holland, 1982).
    \item S. Coleman, \textit{Aspects of Symmetry: Selected Erice Lectures} (Cambridge University Press, 1985).
    \item T. Jacobson, ``Thermodynamics of Spacetime: The Einstein Equation of State,'' \textit{Physical Review Letters} \textbf{75}, 1260 (1995).
    \item G. E. Volovik, \textit{The Universe in a Helium Droplet} (Oxford University Press, 2003).
    \item E. Verlinde, ``On the Origin of Gravity and the Laws of Newton,'' \textit{Journal of High Energy Physics} \textbf{1101}, 029 (2011).
\end{enumerate}

\end{document}